\documentclass[pra,twocolumn,aps]{revtex4-1}
\usepackage{bm}%
\usepackage{float}
\expandafter\ifx\csname package@font\endcsname\relax\else
 \expandafter\expandafter
 \expandafter\usepackage
 \expandafter\expandafter
 \expandafter{\csname package@font\endcsname}%
\fi

\usepackage{array}
\usepackage{amsmath,amssymb}
\usepackage{MnSymbol}
 \usepackage{tikz-feynman} 
\usetikzlibrary{plotmarks}
\usepackage{graphicx}
\usepackage{mathrsfs}
\usepackage{bm}
\usepackage{mathtools}
\usepackage{epstopdf}
\usepackage{hyperref}

\hypersetup{%
   pdfpagemode=None, %FullScreen,
   pdfstartpage=1,
   pdfmenubar=true,
   pdftoolbar=true,
   colorlinks = true,
   linkcolor=blue,
   citecolor=blue,
   urlcolor=blue,
   bookmarksopen=false
 }

 \usepackage{amsmath}
 \usepackage[T1]{fontenc}
\usepackage{mwe}    % loads �blindtext� and �graphicx�
\usepackage{subfig}
\usepackage{amsfonts}
\usepackage{mathtools}
\usepackage{graphicx}
\usepackage{fixme}
\usepackage{color}

\begin{document}
\title{Polariton dynamics in strongly interacting quantum many-body systems }
\author{A.\ Camacho-Guardian$^1$} 
\author{K.\ Knakkergaard Nielsen$^1$}
\author{T.\ Pohl$^1$}
\author{G.\ M.\ Bruun$^{1,2}$} 
\affiliation{$^1$Department of Physics and Astronomy, Aarhus University, Ny Munkegade, 8000 Aarhus C, Denmark}
\affiliation{$^2$Shenzhen Institute for Quantum Science and Engineering and Department of Physics, Southern University of Science and Technology, Shenzhen 518055, China}
\begin{abstract}
We develop a theory for light propagating in an atomic Bose-Einstein condensate in the presence of strong
 interactions. The resulting many-body correlations are shown to  have profound effects on the optical properties of this interacting medium. 
  For weak  atom-light coupling, there is   a well-defined quasiparticle, the polaron-polariton, supporting light propagation with spectral features differing significantly 
  from the non-interacting case. The damping of the polaron-polariton depends non-monotonically on the  light-matter coupling strength, initially increasing and then 
 decreasing. This gives rise to an interesting cross-over between two quasiparticles: a bare polariton and a polaron-polariton, separated by a complex and lossy mixture of light and matter.

\end{abstract}
\date{\today}
\maketitle

\section{Introduction}
The ability to prepare, control, and probe cold matter systems via external light fields is at the heart of modern developments in  atomic physics, quantum optics, many-body physics, 
and quantum technologies. Here, electromagnetically induced transparency (EIT) presents a particularly powerful approach to achieve strong light-matter coupling at greatly reduced losses. This effect opens up numerous applications, from cooling~\cite{Lechner2016} and trapping~\cite{Wang2018} techniques, to the realization of quantum memories~\cite{Hsiao2018} and ultraslow propagation 
of light in the form of dark-state polaritons~\cite{Fleischhauer2000}.   EIT 
has been observed in a wide variety of media including hot atomic vapors~\cite{Phillips2001}, cold atomic gases~\cite{Hau1999,Schnorrberger2009,Zhang2009}, 
Rydberg  gases~\cite{Pritchard2010}, and solids~\cite{Longdell2005}. While many of these applications utilize relatively simple optical media, coupling photons to strongly interacting quantum many-body systems would open the door to quantum nonlinear optics, based on their rich spectrum of strong-correlation phenomena~\cite{Chin2010-Feshbach,Carussotto2013}. Indeed, understanding light propagation in strongly correlated environments remains a problem of great scientific and technological significance that is currently attracting increasing interest in both atomic~\cite{Grusdt2016, Thompson2017,Busche2017,Stiesdal2018} and solid-state~\cite{Takemura2014,Sidler2016} settings. Although the promise of combining EIT and strong particle interactions is widely recognized~\cite{Maser2016}, the effects of  environmental coupling on the dynamics of single slow-light quanta remain to be understood.
\begin{figure}[!ht]
\begin{center}
\includegraphics[width=.6\columnwidth]{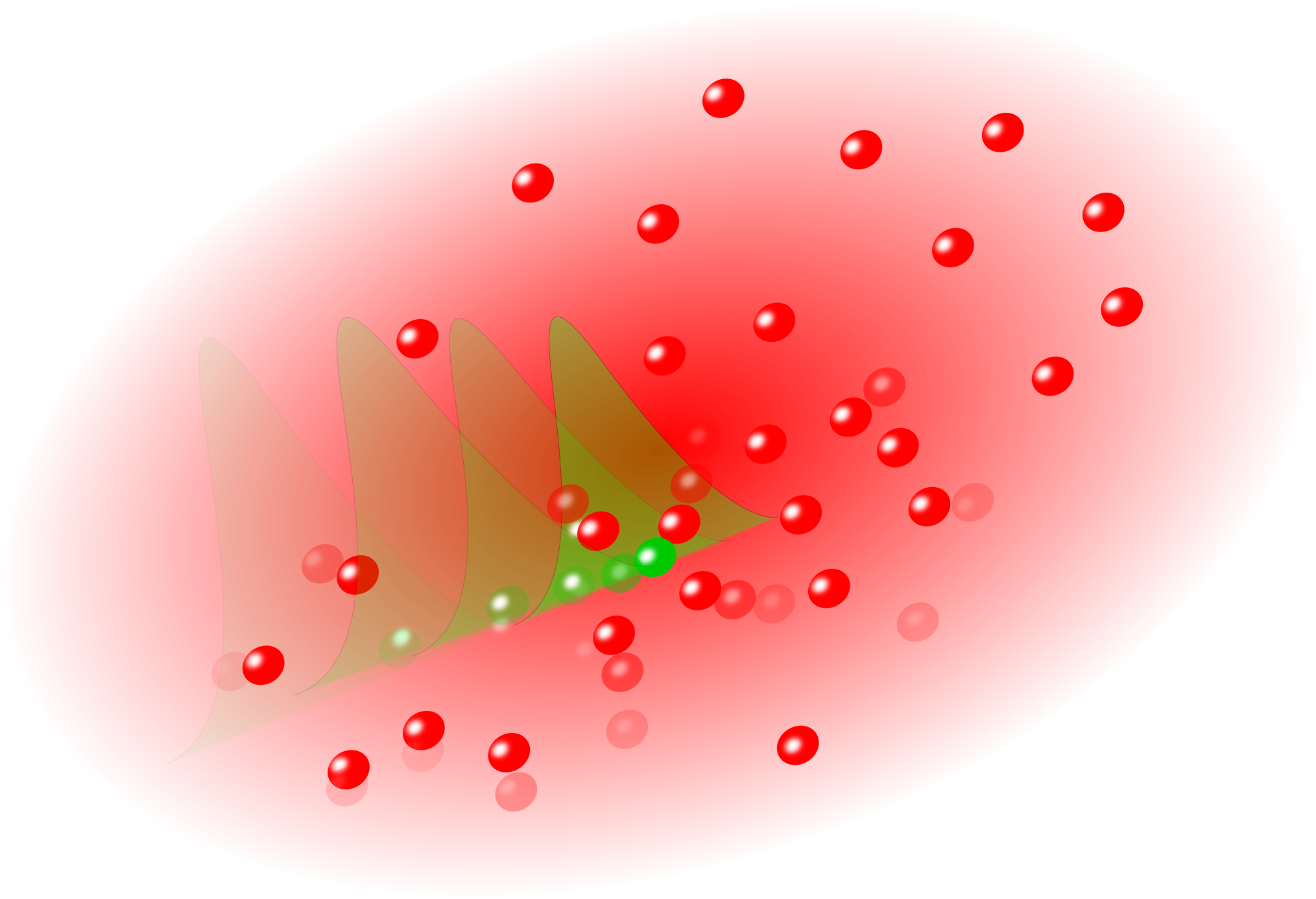} 
\end{center}
\caption{Light propagating as a polaron-polariton in a BEC.}
   \label{Scheme1} 
\end{figure} 
%
%When light propagates through an interacting system its polaritonic spin wave component will typically interact and develop correlations with its surrounding. These correlations can give rise to the formation of quasiparticles, such as polarons. A natural question then arises concerning whether a combined 
%quasiparticle, the polaron-polariton, emerges when both light coupling and interactions are present.  
%While the formation of polarons have been explored in  great detail in cold atomic gases~\cite{Schirotzek09, Koschorreck12, Kohstall12, Scazza17,Jorgensen2016,Hu2016,Yan2019}, and polaron-polaritons 
%have been considered in two-dimensional (2D) semiconductors and atomic systems inside microcavities~\cite{Takemura2014,Sidler2016,Grusdt2016}, the 
%effects of strong interactions on slow light propagation have not been explored  so far. 
Here, we address this problem by developing a non-perturbative theory for the quantum dynamics of a dark-state polariton in Bose-Einstein condensates (BECs), and explore the effects of strong interactions between its spin-wave component and the surrounding condensate.  Interestingly, these interactions lead to the formation of polaron quasiparticles~\cite{Jorgensen2016,Hu2016}, i.e. phonon-dressed impurities, while the formation of dark-state polaritons corresponds to photon-dressing of the same impurity state. The developed theory makes it possible to explore the competition between the creation of these two quasiparticles. While this generally causes the formation of complex light-matter states with substantial environmental dissipation, we identify regimes in which light propagation can be understood in terms of a well-defined a quasiparticle that features reduced decoherence and inherits the properties of both quasiparticle excitations. This polaron-polariton state is shown to have a narrowed EIT linewidth and an even lower group velocity compared to the bare slow-light polariton. Our study establishes a general theoretical framework for quantum optics in strongly interacting systems and will provide a  guide for achieving coherent interfacing and photon-photon interactions in atomic gases and semiconductor materials.

%
%
%Here, we address this important problem and develop a non-perturbative theory for light propagation under EIT conditions in the presence of strong many-body correlations 
%between the photons and a surrounding Bose-Einstein condensate (BEC). Despite the fact that the underlying dark-state is almost entirely composed of the impurity excitation~\cite{Fleischhauer2000}, which 
%forms a polaron in the absence of light, we find that its propagation dynamics cannot
%in general  be described in terms of an effective polaron-polariton theory. Yet, we identify a parameter regime in which light propagation can be described in terms of polaron-polaritons, where pronounced absorption minima persists and feature large shifts that are indicative of the underlying polaron. 
%The emerging combined quasiparticle is found to have a narrowed EIT linewidth and reduced group velocity compared to the bare slow-light polariton. 
%In other regimes, the interplay between light and interactions leads to a complex quantum state with no well-defined quasiparticle, whereas the non-interacting 
%dark-state polariton surprisingly is found to emerge for strong light-matter coupling. We also identify other distinct many-body imprints on the light propagation such as 
%continuum of states causing an enhanced transmission away from the EIT resonance. 

\section{Model}
We consider atoms of mass $m$ with three internal states $|b\rangle$, $|e\rangle$, and 
$|c\rangle$. A quantised probe beam  couples the $|b\rangle$ and  $|e\rangle$ states with a single-photon coupling $g$, whereas a 
classical control field couples the $|e\rangle$ and $|c\rangle$ states with Rabi frequency $\Omega$, forming a 
so-called $\Lambda$-scheme. Within the rotating wave approximation, the Hamiltonian can be written as
\begin{gather}
H=\sum_{\mathbf p}[\epsilon_{\mathbf p}b^\dagger_{\mathbf p}b_{\mathbf p}+\epsilon^{(e)}_{\mathbf p}e^\dagger_{\mathbf p}e_{\mathbf p}+
\epsilon_{\mathbf p}^{(c)}c^\dagger_{\mathbf p}c_{\mathbf p}+cp\gamma_{\mathbf p}^\dagger\gamma_{\mathbf p}]\nonumber\\
+\sum_{\mathbf p}[\Omega e_{\mathbf p}^\dagger c_{\mathbf p-\mathbf k_{\text{cl}}}+\sum_{\mathbf q}ge^\dagger_{\mathbf p+\mathbf q} b_{\mathbf p}\gamma_{\mathbf q}+\text{h.c.}]+
\nonumber\\
\sum_{\mathbf p,\mathbf p',\mathbf q}[V_\text{B}(q) b^\dagger_{\mathbf p+\mathbf q}b^\dagger_{\mathbf p'-\mathbf q}b_{\mathbf p'}b_{\mathbf p}/2
+V(q)b^\dagger_{\mathbf p+\mathbf q}c^\dagger_{\mathbf p'-\mathbf q}c_{\mathbf p'}b_{\mathbf p}]
\label{Hamiltonian}
\end{gather} 
where the operators  $b^\dagger_{\mathbf p}$, $c^\dagger_{\mathbf p}$, and $e^\dagger_{\mathbf p}$ create an atom with momentum $\mathbf p$ and 
kinetic energy $\epsilon_{\mathbf p}=  p^2/2m$ in the atomic state $|b\rangle$, $|c\rangle$, and $|e\rangle$ respectively. 
The atomic states are such that 
 $\epsilon_{\mathbf p}^{(e)}=\epsilon_{\mathbf p}+\epsilon_e$ and $\epsilon_{\mathbf p}^{(c)}=\epsilon_{\mathbf p}+\epsilon_c+\omega_\text{cl}$ with $\epsilon_{e/c}$ their bare state energies respectively. Here
$\epsilon_{\mathbf p}^{(e)}$  includes  the Lamb shift due to the coupling $g$ to the $|b\rangle\otimes|\gamma\rangle$ continuum.
  The operator  $\gamma^\dagger_{\mathbf p}$ creates a photon with momentum 
$\mathbf p$ and kinetic energy $cp$ with $c$ the speed of light in a vacuum. The second line of Eq.~\eqref{Hamiltonian} describes 
the coupling between the atoms and the  probe photons as well as the classical control field. 
Note that the classical field with wave vector ${\mathbf k}_\text{cl}$ ($\omega_{\text{cl}}=c|{\mathbf k}_\text{cl}|$) decreases the momentum of the $|c\rangle$ atoms by $\mathbf k_{\text{cl}}$ 
compared to the $|b\rangle$ and $|e\rangle$ atoms. 
The interaction $V_\text{B}(q)=4\pi a_B/m$ describes the interaction between two atoms in state $|b\rangle$, and $V(q)=\mathcal T_\nu=4\pi a/m$ denotes the interaction between a 
$|b\rangle$- and a $|c\rangle$-state atom. Both interactions are short range and accurately  characterised by the scattering lengths $a_\text{B}$ and $a$ respectively. 
We use units where the system volume and $\hbar$ are both one.  
The $|b\rangle$-atoms form a weakly interacting 3D BEC with density $n=k_n^3/6\pi^2$ and  $0<k_na_\text{B}\ll 1$. 
The  excitation spectrum  of the BEC is  given by Bogoliubov theory, i.e.\ 
$E_\mathbf{p}=\sqrt{\epsilon_{\mathbf p}(\epsilon_{\mathbf p}+2\mu_B)}$ with $\mu_B=4\pi a_\text{B}n/m$ its chemical potential. 

As illustrated in Fig.~\ref{Scheme1}, we  consider a photon with momentum $\mathbf k$ propagating inside the BEC. This excites an 
atom out of the BEC and into the $|e\rangle$- and $|c\rangle$-states via the $\Lambda$-scheme. We focus on the case of a small density 
of $|c\rangle$ atoms so that they can be regarded as impurities  in  the BEC. Strictly speaking, this corresponds to the limit of a single 
photon propagating through the BEC, but in analogy with the case of impurities in atomic gases in the absence of light~\cite{Massignan2014,Jorgensen2016, Hu2016}, we expect this picture to be accurate as long as the density of the  $|c\rangle$ atoms is much smaller than that of the BEC, which 
consequently acts as a particle reservoir. This furthermore means that we can ignore   $|e\rangle-|e\rangle$, $|e\rangle-|c\rangle$, and $|c\rangle-|c\rangle$ 
interactions since the densities of these states are so low. Finally, the scattering length $a$ describing the $|b\rangle-|c\rangle$ interaction is taken to be tuneable 
so that the unitary regime $k_n|a|\gtrsim 1$ of strong interaction can be reached.

\section{Discussion}
Before we plunge into  detailed calculations, let us discuss the main physical concepts and results. 
The system combines two paradigmatic quasiparticles, the dark-state polariton giving rise to EIT in absence of atomic interactions, and the 
Bose polaron emerging due to interactions when there is no light. The polariton wave function is 
$|D_{\mathbf k}\rangle=-\cos\theta\gamma_{\mathbf k}^\dagger|\text{BEC}\rangle+\sin\theta c^\dagger_{\mathbf k-\mathbf k_{\text{cl}}}|\text{BEC}\rangle$
with $|\text{BEC}\rangle$ the wave function of the BEC of $|b\rangle$-atoms and $\cos^2\theta=1/(1+g^2n/|\Omega|^2)$~\cite{Fleischhauer2000,Fleischhauer2005}.
The polaron wave function can on the other hand be written as  
$|\psi_{P,\mathbf k-\mathbf k_{\text{cl}}}\rangle=(\sqrt{Z_P}c_{\mathbf k-\mathbf k_{\text{cl}}}^\dagger+\sum_{\mathbf q}\psi_qc_{-\mathbf q+\mathbf k-\mathbf k_{\text{cl}}}^\dagger\beta_{\mathbf q}^\dagger)|\text{BEC}\rangle$, which describes the 
  impurity  dressed by Bogoliubov excitations created by $\beta^\dagger_{\mathbf q}$~\cite{Li2014,Levinsen2015,Shchadilova2016}. 
 We have introduced the quasiparticle residue $Z_P$  of the polaron and $\psi_q$ are expansion coefficients. 
  Now, it is tempting to assume that the presence of both strong light coupling and atomic interactions would lead to the formation of a 
polaron-polariton of the form
 \begin{align}
| D^P_{\mathbf k}\rangle=-\cos\theta\gamma_{\mathbf k}^\dagger|\text{BEC}\rangle+\sin\theta|\psi_{P,\mathbf k-\mathbf k_{\text{cl}}}\rangle,
\label{PolaronpolaritonWavefn}
\end{align} 
which is a quasiparticle   encompassing simultaneously the polariton and polaron features by replacing the non-interacting impurity
  $c^\dagger_{\mathbf k-\mathbf k_{\text{cl}}}|\text{BEC}\rangle$ state by the polaron  $|\psi_{P,\mathbf k-\mathbf k_{\text{cl}}}\rangle$. We shall  show 
  that although this is an accurate description in certain regimes, it breaks down in other regimes in favour of a complex light-matter quantum state.   
\begin{figure}[!ht]
\begin{center}
\includegraphics[width=1\columnwidth]{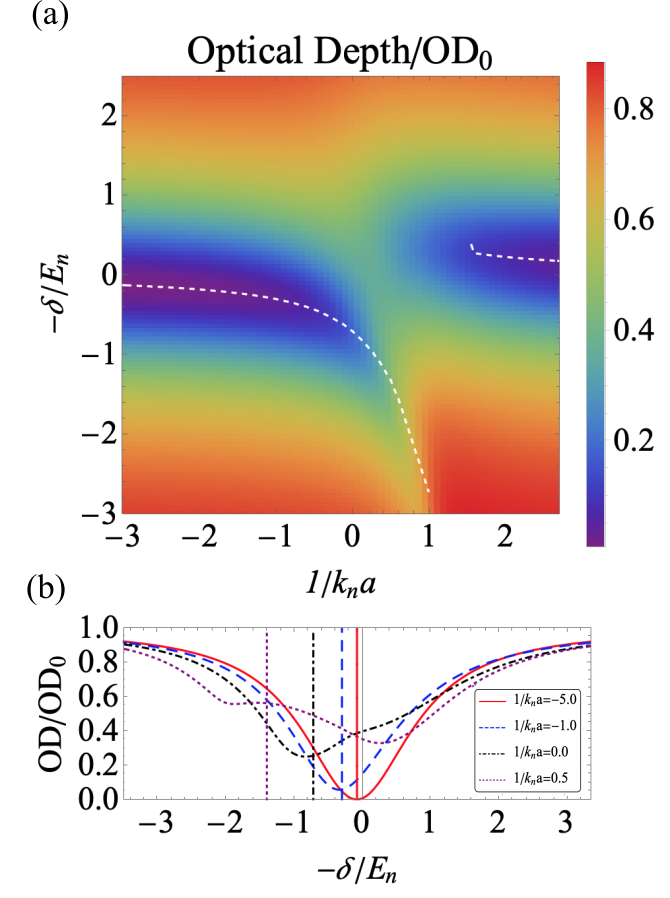} 
\end{center}
\caption{(a) Optical Depth as a function of the two-photon detuning and the atomic interactions. (b) Cross sections of the optical depth for several values 
of the inverse of the interaction strength. The vertical lines give  the polaron energy in absence of  light.}
   \label{Scheme} 
\end{figure} 

Figure \ref{Scheme}(a) summarises our results. It shows the optical depth of a BEC as a function of the two photon detuning  $\delta=\epsilon_c+\omega_\text{cl}-ck$ and the interaction strength $1/k_na$. First, we see a 
pronounced minimum in the optical density, which for weak interactions is almost zero. The minimum is located when the incoming photon energy matches the 
polaron energy (dashed line), reflecting that  the EIT is caused by the formation of a  polaron-polariton  with a small damping.
Also, the width of the EIT minimum narrows with increasing interaction, which is caused by a decreasing polaron residue $Z_P$. The reason is that $Z_P$ by definition determines the overlap between the polaron state $|\psi_{P,\mathbf k-\mathbf k_{\text{cl}}}\rangle$ and the plane wave $|c\rangle$ state. These effects 
reflect the difference between the polaron-polariton and the polariton, which would simply give rise to perfect EIT at the horizontal line 
$\delta=0$. For stronger interactions however, Fig.~\ref{Scheme}(a) shows that the optical depth at the minimum is 
\emph{non-zero}, and that the minimum position is shifted \emph{away} from the polaron energy. This reflects that the polaron-polariton picture has broken 
down and that 
the interplay between light coupling and strong interactions produces a complex state with no well-defined quasiparticle. 
In the rest of the manuscript, we will derive and discuss in detail these as well as other intriguing results  showing the imprints of many-body physics on light transmission.

\section{Field theory}
\begin{figure}[!h]
\begin{center}
\includegraphics[width=1\columnwidth]{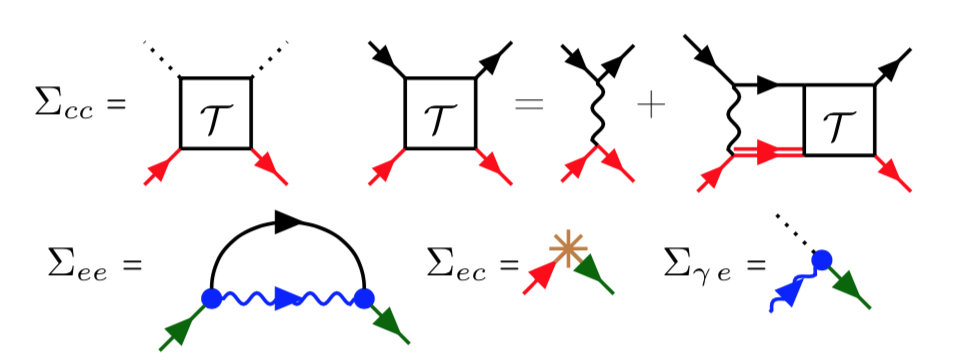}  
\end{center}
\caption{ Feynman diagrams representing our theory for light propagation in the presence of strong interactions. The self-energy for the $|c\rangle$ atoms describing strong $|b\rangle-|c\rangle$ interactions leading to polaron formation in the absence of light is shown in the top panel. The coupling to the classical and quantum light  responsible for EIT in absence of interactions
is shown in the bottom panel.      
        }
        
        \label{Diagrams}
        \end{figure}

In order to develop a non-perturbative theory that can simultaneously account for strong-light matter coupling as well as atomic interactions, we introduce the imaginary time Green's function ${\mathcal G}({\mathbf p},\tau)=-\langle T_\tau\{\Psi_{\mathbf p}(\tau)\Psi_{\mathbf p}^\dagger(0)\}\rangle$, where 
$T_\tau$ denotes time ordering and 
${\Psi}_{\mathbf p}=[\gamma_{\mathbf p},e_{\mathbf p},c_{\mathbf p-\mathbf k_{\text{cl}}}]^T$. 
Due to the coupling between light and atoms, the  Green's function  is a $3\times3$ matrix and we write in frequency space   ${\mathcal G}^{-1}(\mathbf p,z)={\mathcal G}^{(0)}(\mathbf p,z)^{-1}-\Sigma(\mathbf p,z)$ as
\begin{align}
{\mathcal G}^{-1}(\mathbf p,z)=
\begin{bmatrix}
z-cp&-g\sqrt n&0\\
-g\sqrt n&z-\epsilon^{(e)}_{\mathbf p}-\Sigma_{ee}&-\Omega\\
0&- \Omega &z-\epsilon^{(c)}_{\mathbf p-\mathbf k_{\text{cl}}}-\Sigma_{cc}
\end{bmatrix}.
\label{eq:GreensFn}
\end{align} 
The  off-diagonal self-energies $\Sigma_{ce}=\Sigma_{ec}=\Omega\hspace{0.5cm}$ and $\Sigma_{\gamma e}=\Sigma_{e \gamma}=g\sqrt{n}$ (both real)
give the light-matter couplings responsible for the EIT phenomena~\cite{Fleischhauer2000},
which can be read off from the second-line of Eq.~\ref{Hamiltonian}. The diagonal self-energy 
$\Sigma_{ee}$  gives  the decay of the $|e\rangle$-atom due to the coupling $g$ 
to the $|\gamma\rangle\otimes|b\rangle$ continuum described within 
  Weisskopf-Wigner theory~\cite{Weisskopf1930,scully_zubairy_1997,footnote}. Finally, $\Sigma_{cc}=n{\mathcal T}$   describes the scattering 
of a $|b\rangle$-atom out of the condensate by a  $|c\rangle$-atom, which is the dominant process leading to the formation of the Bose polaron~\cite{Li2014}.  The scattering matrix ${\mathcal T}$ is evaluated in the so-called ladder approximation 
accounting for  repeated boson-impurity scattering. This approximation  includes the two-body $|b\rangle-|c\rangle$ scattering giving rise to strong Feshbach interactions exactly, 
and it has turned out to be surprisingly accurate for impurities in atomic gases even for strong interactions~\cite{Massignan2014,Jorgensen2016, Hu2016}.
The scattering matrix is given by
 ${\mathcal T}$
 \begin{align}
\mathcal{T}(\mathbf p,z)=\frac{\mathcal T_\nu}{1-\mathcal T_\nu\Pi(\mathbf p,z)},
\label{Tmatrix}
\end{align}
where  \begin{gather}
\Pi(\mathbf p,z)=-\sum_{\mathbf k,i\omega_\nu}\mathcal G_{11}(\mathbf k,i\omega_\nu)\mathcal G_{cc}(\mathbf p-\mathbf k,z-i\omega_\nu)
\label{pair}
\end{gather} is the  regularised 
propagator for a pair of 
$|b\rangle$- and $|c\rangle$-atoms in the presence of a BEC~\cite{Rath2013,Christensen2015}. Here, the $|c\rangle$-atom propagator $\mathcal G_{cc}$ is given by 
\begin{widetext}
\begin{gather}
\mathcal G_{cc}^{-1}(\mathbf p-\mathbf k_{\text{cl}},\omega)=\mathcal G_{cc}^{(0)}(\mathbf p-\mathbf k_{\text{cl}},\omega)^{-1}
- \frac{|\Omega|^2}{\mathcal{G}_{ee}^{(0)}(\mathbf p,\omega)^{-1}-\Sigma_{ee}(\mathbf p,\omega) -ng^2\mathcal {G}^{(0)}_{\gamma\gamma}(\mathbf p,\omega)}.
\label{cpropagator}
\end{gather}
\end{widetext}
Since atom-light coupling is crucial for EIT physics, we 
have included the self-energies $\Sigma_{ce}$ and $\Sigma_{\gamma e}$ in the impurity propagator in Eq.~\eqref{cpropagator}, 
which describe  the coupling to the excited state $|e\rangle$ and the photon $|\gamma\rangle$.  This
  goes beyond the usual ladder approximation based on bare $c$-propagators or the equivalent variational Chevy ansatz, and it has important 
 consequences as will be discussed in detail below. In Fig.~\ref{Diagrams}  we illustrate the diagrams  corresponding Eq.~\eqref{eq:GreensFn}. A dashed line is a 
 $|b\rangle$-atom emitted from or absorbed into the BEC, a red line is the $|c\rangle$-propagator, a  green line is the $|e\rangle$-propagator, a    wavy blue line  is the photon propagator,  a black line is a  $|b\rangle$-propagator, and a double solid red line corresponds to the impurity propagator including light-matter coupling. 
    The classical field $\Omega$ is indicated  by a $*$,  a $\bullet$ is the dipole matrix element $g$ between the photons and the atoms, and a 
     wavy black line is the $|b\rangle$-$|c\rangle$ interaction.

  After diagonalising Eq.~\eqref{eq:GreensFn}, we obtain the photon Green's function describing light propagation in the BEC. Writing this as
   $\mathcal G^{-1}_{\gamma\gamma}(\mathbf k,\omega)=\epsilon\omega-ck$ allows us to relate the dielectric function $\epsilon=1+\chi$ and the optical susceptibility to the self-energies as~\cite{Mahan2000Book}
\begin{align}
\chi(\mathbf k,\omega) =-\frac{1}{ck}\frac{ng^2}{\omega - \epsilon^{(e)}_{\mathbf k} + i\Gamma_{ee} - \Omega^2\mathcal G_P(\mathbf k-\mathbf k_{\text{cl}},\omega)}.
\label{PhotonSelfEnergy}
\end{align}
Here $\Gamma_{ee}=-\text{Im}\Sigma_{ee}$ gives the decay rate of the excited state,  and 
$\mathcal G^{-1}_P={{\mathcal G}_{cc}^{(0)}}^{-1}-\Sigma_{cc}$, see Appendix~\ref{GreenApen}.

\section{Polaron-polaritons}
We can now show how the polaron-polariton emerges by ignoring the effects of light on the $|b\rangle-|c\rangle$ scattering described by the ${\mathcal T}$-matrix.
As we shall see, this approximation is valid for $\Omega^2/\Gamma_{ee}\ll E_n$ where $E_n=k_n^2/2m$ sets the many-body energy scale. 
Then the term $\mathcal G_P$ in Eq.~\eqref{PhotonSelfEnergy} becomes identical to the polaron Green's function  in the ladder approximation~\cite{Rath2013},
which has a pole at  the undamped polaron ground state energy $E^{(P)}_{\mathbf k+\mathbf q-\mathbf k_{\text{cl}}}+\delta$. 
Close to this pole, we can write
$\mathcal G_{P}(\mathbf k+\mathbf q - \mathbf k_{\text{cl}},ck+\omega)\simeq Z_P/(\omega-E^{(P)}_{\mathbf k+\mathbf q-\mathbf k_{\text{cl}}}-\delta)$,
where $Z_P$ is the quasiparticle residue of the polaron wave function  $|\psi_{P}\rangle$ introduced above. 
 It follows from Eq.~\eqref{PhotonSelfEnergy} that the on-shell susceptibility $\chi(\mathbf k,ck)$ vanishes at this pole, i.e.\ when  $\mathcal G_{P}^{-1}(\mathbf k-\mathbf k_{\text{cl}},ck)=0$. Physically, this  means that the photon can propagate undamped under perfect EIT conditions
 when its   energy   $ck$ matches that of a  polaron with momentum $\mathbf k-\mathbf k_{\text{cl}}$, i.e.
 when $ck=E^{(P)}_{\mathbf k-\mathbf k_{\text{cl}}}+\epsilon_c+\omega_{\text{cl}}$.  In the wave function picture introduced above, this corresponds to light propagation 
  carried by the polaron-polariton state $| D^P_{\mathbf k}\rangle$ given by Eq.~\eqref{PolaronpolaritonWavefn}, 
  instead of the non-interacting polariton $|D_{\mathbf k}\rangle$.
 
To further explore many-body effects on the EIT spectrum, we use a 
 pole expansion of $\mathcal G_P$ in   Eq.~\eqref{PhotonSelfEnergy}. This gives 
\begin{align}
\mathcal G_{\gamma\gamma}(\mathbf k+\mathbf q,ck+\omega)\simeq 
\frac Z{\omega-v_g q - \tilde\delta + i \frac{(\omega-\tilde\delta)^2}{\sigma}},
\end{align}
for the photon propagator around 
the  EIT condition $\delta=-E^{(P)}_{\mathbf k-\mathbf k_{\text{cl}}}$ to 
 first order in the deviations  $\omega$, $\tilde\delta=\delta+E^{(P)}_{\mathbf k-\mathbf k_{\text{cl}}}$, and ${\mathbf q}$, which is taken to be parallel to ${\mathbf k}$
 for simplicity. We have neglected terms involving $\nabla_{\mathbf k}E^{(P)}_{\mathbf k-\mathbf k_{\text{cl}}}\lesssim c_s\ll c$ and defined
\begin{align}
Z&=\frac1{1+g^2n/\Omega_P^2}, &v_g&=Zc,\nonumber\\
 \sigma &= \frac{\Omega_P^2}{\Gamma_{ee}} , &\Omega_P^2&=Z_P\Omega^2.
\label{groupvelocity}
\end{align}
Here,  $Z=\cos^2\theta$ is the residue of the EIT pole in the photon propagator,  which in the wave function formulation 
is simply given by the photon component of the polaron-polariton 
state $| D^P_{\mathbf k}\rangle$ given by Eq.~\eqref{PolaronpolaritonWavefn}. Also,  $v_g$ is 
the group velocity of light, $\sigma$ is the width  of the EIT window, and  $\Omega_P$ is the Rabi frequency renormalised by many-body correlations. From Eq.~\eqref{groupvelocity}, we see in addition to moving the  condition for EIT away from 
$\delta=0$, the formation of the polaron  decreases both the group velocity of light in the BEC and the width of the EIT window through its residue $Z_P<1$.

We return to  Fig.~\ref{Scheme} showing the optical depth of a BEC of length $L$  as   a function of the $|b\rangle$-$|c\rangle$ scattering length $a$ 
and the detuning $\delta$. The transmission is described by the  optical depth 
\begin{align}
 \text{OD}=\frac{\Gamma_{\gamma}kL}{v_g}=\text{Im}\chi kL \text{ where } \Gamma_{\gamma}=Zc\text{Im}\chi
 \end{align} 
 is the damping rate of the photons. The  optical depth $\text{OD}_0=ng^2L/\Gamma_{ee}c$ in the absence of the classical control field serves as a reference. To demonstrate that the physics we discuss is within experimental reach, 
 we consider the $ 4^2\text{S}_{1/2} $ to $ 4^2\text{P}_{1/2}$ transition in 
 $^{39}\text{K}$, which   has already been employed in recent  EIT and polaron experiments~\cite{Jorgensen2016, Lampis2016}. 
 For this  transition, $\Gamma_{ee}=\pi\times\,2.978\text{MHz}$ corresponding to a wavelength $\lambda=2\pi/k\simeq 700.1\text{nm}$.
 Taking a typical BEC density of $n= 2\times 10^{14}\text{cm}^{-3}$, this gives   $E_n=k_n^2/2m\simeq 420\text{kHz}$, and 
 using $g^2=3\pi\,c\Gamma_{ee}/k^2$ from  Weisskopf-Wigner theory yields $\sqrt{n}g\simeq 6.1\times 10^{5}E_n$.  In order to 
 resolve many-body physics in the  spectrum, we choose a classical light coupling $\Omega$ so that the  width $\sigma \simeq \Omega^2/\Gamma_{ee} = 518 \text{kHz}$ is comparable to $E_n$. Finally,   the impurity momentum ${\mathbf k-\mathbf k_{\text{cl}}}$, the temperature and the
 one-photon detuning $\Delta=\varepsilon^{(e)}_{\mathbf k}-ck$ are all zero. 
We also plot as dashed lines in Fig.~\ref{Scheme}(a) the attractive and repulsive polaron energies in the absence of light, determined by the pole of $\mathcal G_P$ 
with no light, i.e.\ $\Omega=0$. Figure \ref{Scheme}(a) clearly demonstrates that the optical depth essentially vanishes when the two-photon detuning matches the polaron energy, $-\delta=E^{(P)}_{\mathbf k-\mathbf k_{\text{cl}}}$,   for weak attractive coupling $1/k_na\lesssim -1$. 
 This corresponds to  the formation of a polaron-polariton leading to  EIT as described above. 
 
 Figure \ref{Scheme}(b) shows vertical cuts  for several values of the interaction strength and 
 the vertical lines correspond to the polaron energy in absence of any  light.  We see that 
  the  optical depth at the EIT resonance is in general larger for $k_na>0$ compared to the attractive side, reflecting that the repulsive polaron is not the ground state
  so that  it can decay into lower lying states such as the Feshbach molecule  even in the absence of light.
In addition, we see an interesting  double dip structure in the optical depth for strong interactions $0\lesssim1/k_na\lesssim 1$. This is a genuine many-body effect beyond the quasiparticle picture: It is caused by a continuum of states  involving Bogoliubov excitations of the BEC, which increases the transparency of the BEC for detunings away from the polaron energies.

\section{Light induced damping}
From Figs.~\ref{Scheme}(a)-(b),  we  also see that  for stronger interactions, the optical depth  increases at the minimum, 
which moreover is shifted away  from the polaron energy. 
As the interaction $k_n|a|$ increases, the EIT minimum is displaced away from the polaron energy and the optical depth increases  
becoming substantial at  unitarity  $1/k_na=0$.  This is caused by  the  interplay between the scattering and the light coupling, 
which leads to additional decay and eventual breakdown of the polaron, \emph{even} when it is the ground state in the absence of light.  
The key point is that while the coupling $\Omega$ of the $|c\rangle$-state to the lossy $|e\rangle$-state  is suppressed for the 
EIT resonant momentum ${\mathbf k}-{\mathbf k}_{\text{cl}}$, it can be significant for other momenta where the photon is off-resonant. The remaining 
 light coupling to the  $|e\rangle$-state is controlled by the ratio $\Omega/\Gamma_{ee}$ and leads to damping of the impurity.
     This is of course  irrelevant for EIT physics in the absence of interactions where the impurity momentum is fixed to ${\mathbf k}-{\mathbf k}_{\text{cl}}$ by 
     the incoming light. In the presence of interactions however,  atom scattering changes the momentum of the impurity to values, where the state $|c\rangle$ 
     couples strongly to the lossy $|e\rangle$-state
   and this damping mechanism kicks in. One can show that for $\Omega/\Gamma_{ee}\ll1$, the resulting damping of the
   polaron  with resonant      momentum ${\mathbf k-\mathbf k_{\text{cl}}}=0$ is 
    $\Gamma_P\propto(1-Z_P)\Omega^2/\Gamma_{ee}$ see details in Appendix~\ref{ApenB}. This in turn results in a damping of the photons
     and a corresponding non-zero minimal optical depth given respectively by 
   \begin{align}
\Gamma_\gamma & \simeq \Gamma_P + \frac{\Gamma_P^2\Gamma_{ee}}{|\Omega_P|^2}, &\text{OD}&= \Gamma_\gamma\frac L{v_{g}} \propto \text{OD}_0 (1 - Z_P),
  \label{ODatresonance}
  \end{align} 
 for  $\Omega/\Gamma_{ee}\ll1$, $\Omega^2/\Gamma_{ee}\ll E_n$ and $\sqrt{n}g \gg \Omega_P,$ see Appendix~\ref{ApenB}.
 Equation  \eqref{ODatresonance} relates the optical depth of the medium to the incoherent excitations forming the polaron, which have a spectral weight of $1-Z_P$.
  In this sense, it provides  a profound link between the propagation of light and the quasiparticle properties of the polaron. This, combined 
  with the fact that the position and value at the minimum of the optical depth are determined by the energy and  residue of the polaron respectively,  demonstrates 
  how strong light-matter coupling and  slow-light provides a powerful new platform for probing quantum many-body physics in a non-demolition scheme.

Increasing $\Omega$ further eventually makes the impurity states with momenta different from $\mathbf k-\mathbf k_{\text{cl}}$    so strongly damped,
  that scattering into them is suppressed. As a result, both the energy shift and the damping of the impurity with resonant momentum decreases, and the EIT spectrum  
approaches that  of a \emph{ideal} gas. In other words, interaction effects are suppressed for a strong control field giving rise to a non-monotonic 
dependence of the damping and the eventual re-emergence of the non-interacting 
polariton for large $\Omega$. This surprising effect can only be described using a non-perturbative theory taking into account the repeated scattering of impurities on the BEC, see Appendix~\ref{GreenApen}. 
\begin{figure}[!ht]
\begin{center}
\includegraphics[width=1\columnwidth]{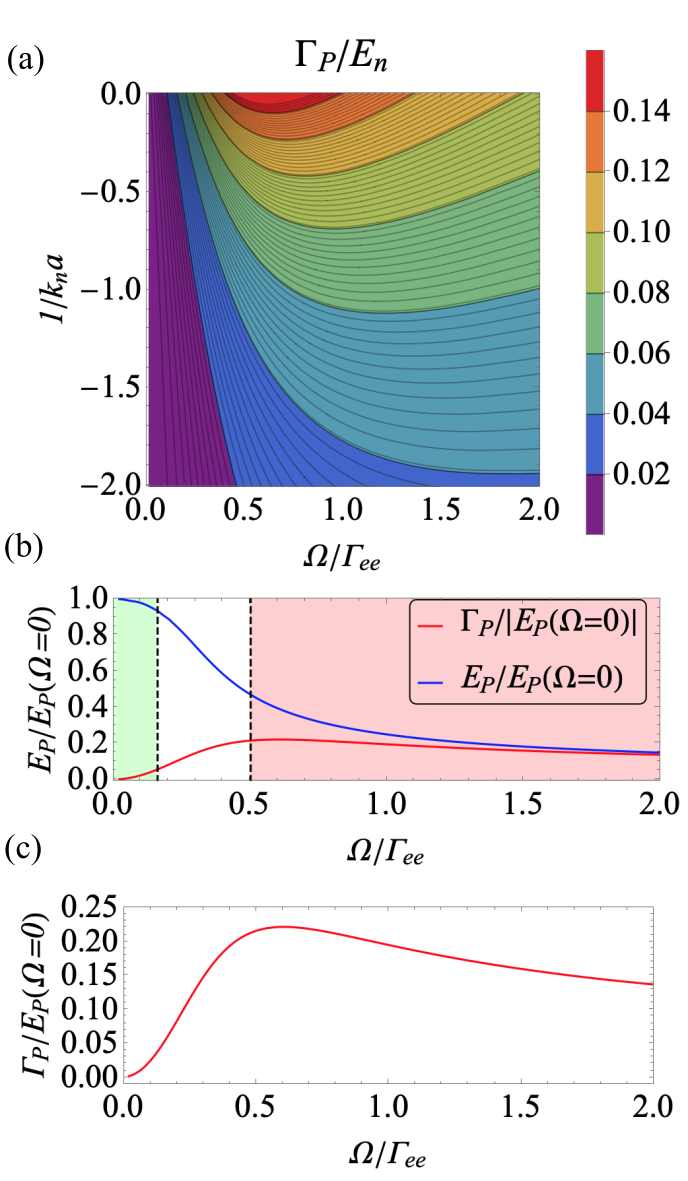} 
\end{center}
\caption{(a) Decay rate of the impurity at the EIT minimum  as a function of $\Omega/\Gamma_{ee}$ and $1/k_na$. (b) Polaron energy and damping 
as a function of  $\Omega/\Gamma_{ee}$  at unitarity $1/k_n a=0.$  (c) The non-monotonicity of the damping is illustrated.}
   \label{FigGamma} 
\end{figure} 

\section{Regimes of light propagation}
The different regimes of light propagation around the EIT minimum in the presence of interactions are shown 
 in Fig.~\ref{FigGamma}(a). It plots  the damping rate of the impurity
  for the detuning $\delta$ giving the minimal optical depth as a function of   the interaction strength $1/k_na$   and the classical light coupling $\Omega/\Gamma_{ee}$,
keeping $\Gamma_{ee}$ fixed. All other parameters are as in Fig.~\ref{Scheme}. In agreement with the discussion above, we see that the damping 
of the impurity depends non-monotonically on $\Omega/\Gamma_{ee}$ for fixed coupling strength $1/k_na$, as shown also in Fig.~\ref{FigGamma}(c).
For $\Omega/\Gamma_{ee}\ll 1$, the damping is small and  light propagates in the form of a well-defined polaron-polariton giving rise to EIT with 
 a small but finite residual absorption.   The damping increases  with increasing $\Omega/\Gamma_{ee}$ and it becomes substantial for strong coupling $1/k_n|a|\gg1$ and intermediate classical atom-light coupling 
$0.3\lesssim\Omega/\Gamma_{ee}\lesssim 1$. Finally, both the decay and the energy shift  of the impurity start to decrease for even stronger light coupling and
 the ideal gas EIT spectrum governed by the non-interacting polariton re-emerges.  Note this re-emergence of ideal gas slow light propagation occurs for arbitrarily large 
  impurity-boson scattering length $a$, since scattering is suppressed into lossy $|c\rangle$-states with  
    off-resonant momentum $\neq {\mathbf k}-{\mathbf k}_{\text{cl}}$.

To illustrate this  in  more detail, we plot in Fig.~\ref{FigGamma}(b) the polaron  energy $E_P$ and its decay
 rate $\Gamma_P$ at unitarity $1/k_na=0$ as a function of  $\Omega/\Gamma_{ee}$, in units of the polaron energy $E_P(\Omega=0)$ in the absence of light.  The damping 
initially  increases as $\Omega^2$ in agreement with the analysis above, and it  is much smaller than the 
 energy shift for $\Omega/\Gamma_{ee}\ll 1$ so that the polaron-polariton is well-defined. For intermediate values  $0.3\lesssim\Omega/\Gamma_{ee}\lesssim 1$
 however,  the damping is of the same order as the  energy, which is  significantly shifted away from the polaron energy in the absence of light. 
   In this region, the quantum state is a complex mixture of light and photons with  no well-defined quasiparticle resulting in significant decay. 
 Finally,      both the energy shift and the damping become small for larger values of $\Omega/\Gamma_{ee}$,  so that the non-interacting 
 dark-state polariton emerges. 
 
\section{Concluding remarks}
We developed a non-perturbative theory for light propagation through a Bose Einstein condensate in the presence of strong interactions, that permits to explore the interplay of particle correlations and strong light-matter coupling. We have shown that the associated competition between the formation of polaritonic and polaronic quasiparticles can be observed and probed directly via the transmission spectrum of the interacting medium.  This includes large deviations from non-interacting EIT as well as  light-induced damping, and it 
offers a powerful non-destructive setup to manipulate and probe many-body physics.  The presented approach can straightforwardly be generalised to include other interaction effects between the atomic states, or to describe other systems such as exciton-polaritons in semiconductors~\cite{Kasprzak2006,Takemura2014,Sidler2016}. It therefore 
provides a powerful framework for describing systems with  light-matter coupling in the presence of strong interactions, and enables future explorations into key problems  such as  generating 
 strong photon nonlinearities by polaron-polaron interactions~\cite{Camacho2018b}.
 
\section{Acknowledgments}
We thank Jan Arlt and Luis Ardila for helpful discussions. A. C.G and G.M B acknowledge financial support from the Villum Foundation and the Independent Research Fund Denmark - Natural Sciences via Grant No. DFF - 8021-00233B. This work was supported by the EU through the H2020-FETOPEN Grant No. 800942640378 (ErBeStA), by the DFG through the SPP1929, and by the DNRF through a Niels Bohr Professorship to T. P.

\appendix
\begin{widetext}
\section{Scattering matrix and Green's functions}
\label{GreenApen}

The Dyson's equation yields for  the excited state
\begin{align}\label{epropagator}
\mathcal {G}_{ee}^{-1}(\mathbf p,\omega)={\mathcal G_{ee}^{(0)}(\mathbf p,\omega)}^{-1}-ng^2\mathcal {G}^{(0)}_{\gamma\gamma}(\mathbf p,\omega) 
-|\Omega|^2\mathcal {G}_{P}(\mathbf p-\mathbf k_{\text{cl}},\omega)-\Sigma_{ee}(\mathbf p,\omega),
\end{align}
where  $\Sigma_{ee}(\mathbf p,\omega)$ accounts for the Lamb shift and the decay of the excited state, while $G_P$ is given
in the main  text. 

Finally, the Dyson's equation for the photon field gives $\mathcal G_{\gamma\gamma}^{-1}(\mathbf k,\omega)=\epsilon(\mathbf k,\omega)\omega-ck,$ and relates the Green's functions and self-energies in Fig.~\ref{Diagrams} to the optical susceptibility $\chi(\mathbf k,\omega)$  via  $\epsilon(\mathbf k,\omega)=1+\chi(\mathbf k,\omega)$~\cite{Mahan2000Book} The expression for   $\chi(\mathbf k,\omega)$ is provided in the main text.

Here, we provide more details concerning the light induced damping of the polaron discussed in the main text. This damping enters via the impurity states with  
momenta differing from $\mathbf k-\mathbf k_{\text{cl}}$ inside the scattering matrix in Eq.~\ref{Tmatrix}. Let the momentum of the impurity inside the 
scattering matrix be $\mathbf p-\mathbf k_{\text{cl}}$ and define $\mathbf q=\mathbf p-\mathbf k$. When 
 $cq\ll  ng^2/\Gamma_{ee}$, it follows from Eq.~\eqref{cpropagator}
 that 
 \begin{gather}
\label{Aprox1}
\mathcal G_{cc}^{-1}(\mathbf q,ck+\omega)\approx \mathcal G_{cc}^{(0)}(\mathbf p-\mathbf k_{\text{cl}},\omega)^{-1}-cq\frac{\Omega^2}{ng^2}\approx
 \mathcal G_{cc}^{(0)}(\mathbf p-\mathbf k_{\text{cl}},\omega)^{-1}-v_gq,
\end{gather} 
where we have used $v_g\simeq c\Omega^2/gn^2$. Equation \eqref{Aprox1} shows that the impurities with momenta close to the resonant momentum $\mathbf k-\mathbf k_{\text{cl}}$
are only weakly 
coupled to the excited state and thus have a long lifetime. The linear dispersion leads to an additional source of decay, Cherenkov radiation.

For  $cq\gg ng^2/\Gamma_{ee}$ on the other hand, we get from Eq.~\eqref{cpropagator}
\begin{gather}
\label{Aprox2}
\mathcal G_{cc}^{-1}(\mathbf q,ck+\omega)\approx \mathcal G_{cc}^{(0)}(\mathbf p-\mathbf k_{\text{cl}},\omega)^{-1}-\frac{\Omega^2}{\omega-\epsilon^{(e)}_{\mathbf k+\mathbf q}+i\Gamma_{ee}}.
\end{gather} 
That is, for momenta far away from the resonant momentum the impurity couples to the excited state, which results in decay. 
%We consider in Eqs. \ref{Aprox1} and \ref{Aprox2} the case of interest $\delta=-E^P$ and $\Delta=0$ for the two-photon and one-photon detuning respectively and for simplicity $\mathbf k-\mathbf k_{\text{cl}}=0$ and $\mathbf q$ parallel to $\mathbf k$.  

To estimate how this decay of impurities with momenta different from $\mathbf k-\mathbf k_{\text{cl}}$ gives rise to a decay of the polaron with resonant momentum 
$\mathbf k-\mathbf k_{\text{cl}}$ via the interaction, we first consider the case  
 $\Omega/\Gamma_{ee}\ll 1$ and $\Omega^2/\Gamma_{ee}\ll E_n$.  Estimating the propagators inside the scattering matrix to be given by Eq.~\eqref{Aprox2} then 
 yields 
   \begin{gather}
 \Gamma_P\simeq-Z_P\text{Im}\Sigma_P(E^{(P)}_{\mathbf k-\mathbf k_{\text{cl}}}+i\Omega^2/\Gamma_{ee})\approx (1-Z_P)\Omega^2/\Gamma_{ee}.
 \label{damp}
 \end{gather}

In the opposite regime where $\Omega^2/\Gamma_{ee}\gg E_P$, the pair-propagator  in Eq.~\eqref{Tmatrix} can be approximated by $\Pi(\mathbf p,\omega)\propto -im^{3/2}\sqrt{\omega+i\Omega^2/\Gamma_{ee}}$. For $\Omega^2/\Gamma_{ee}$ larger that the typical atomic energies, 
this  suppresses the boson-impurity scattering matrix in in Eq.~\eqref{Tmatrix} and thereby the impurity self-energy. Thus, one recovers the non-interacting 
dark state polariton for large $\Omega/\Gamma_{ee}$. 
 
 To illustrate the imprints of the light on the atomic scattering, we neglect those in Fig.~\ref{FigC} (left), and compare to the physical case discussed in the main text. For illustration purposes, we show the latter in Fig.~\ref{FigC} (right), which fully includes the light-matter coupling.  In Fig.~\ref{FigC} (left) the optical depth at resonance $\delta=-E_P$ is strictly zero, illustrating that the ground-state polaron in absence of any light-matter coupling is undamped. In agreement with the theory, the width of the EIT reduces, as a consequence of the normalised Rabi frequency $\Omega^2_P= Z_P|\Omega|^2$  that decreases with the residue of the polaron $Z_P$.
 \begin{figure}[!h]
\begin{center}
\includegraphics[width=0.9\columnwidth]{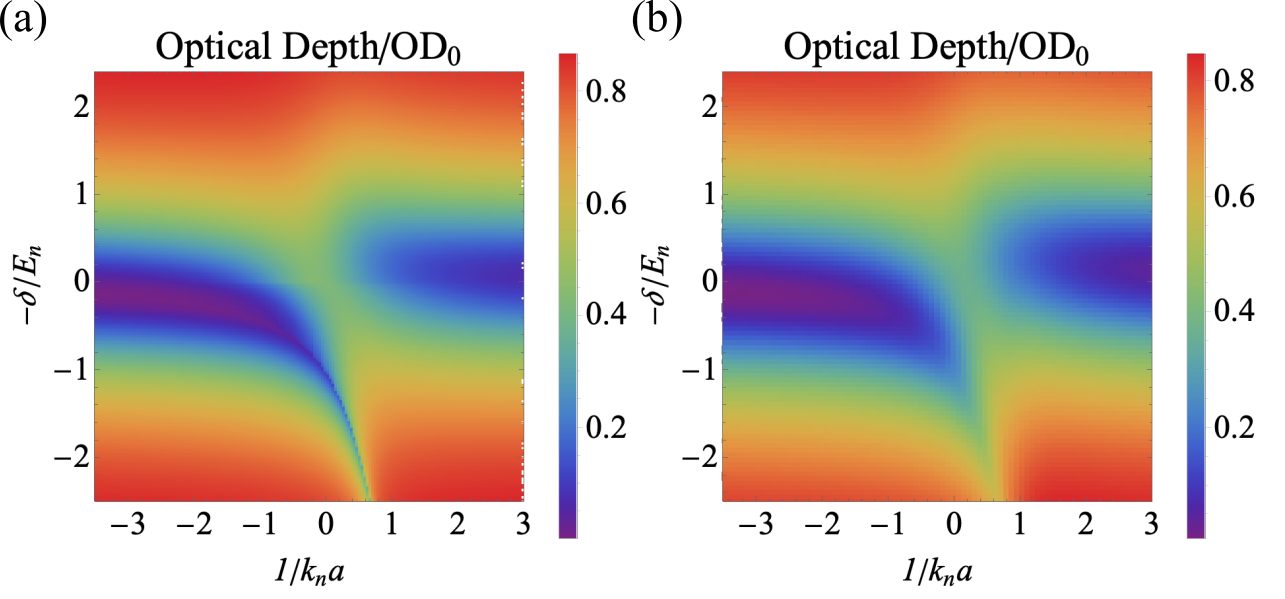}  
\end{center}
\caption{(Left) Idealised picture of an undamped polaron-polariton. (Right) Physical polaron-polariton, here the light-matter coupling modifies the atomic scattering leading to deviations from the idealised undamped polaron-polariton picture
}
   \label{FigC} 
\end{figure} 
 The idealised undamped polaron-polariton corresponds to $ck_n\gg ng^2/\Gamma_{ee}$ and $\Omega^2/\Gamma_{ee}\ll E_n$ where the scattered $|c\rangle$-states are effectively decoupled from the resonant photons and the classical control field. In this limit, the atomic interactions can be described by the scattering matrix in absence of any light-coupling~\cite{Rath2013}.
 
  \section{Damping of the polaron-polariton}
  \label{ApenB}
 The damping of the polaron in turn gives rise to a damping of the polaron-polariton  given by
 \begin{gather}
 \Gamma_{\gamma}=\frac{\tilde Z ng^2 \Gamma_P}{\Gamma_P\Gamma_{ee}+|\Omega_P|^2},
 \label{decayph}
 \end{gather}
 where
 \begin{gather}
 \tilde Z=\frac{1}{1+\frac{ng^2(|\Omega_P|^2+\Gamma_P^2)}{(|\Omega_P|^2+\Gamma_{ee}\Gamma_P)^2}},
 \end{gather}
 is the modified residue of the EIT pole due to the light coupling. For $|\Omega_P|^2\ll \Gamma_P$ and taking $ng^2\gg |\Omega_P|^2,$  the decay of the photon is $\Gamma_\gamma=\Gamma_P\left(1+\Gamma_P\Gamma_{ee}/|\Omega_P|^2\right).$ The optical depth $\text{OD}\propto \text{OD}_0(1-Z_P)$ can be obtained by using 
 Eq.~\eqref{damp} in  Eq.~\eqref{decayph}.
 \end{widetext}
\bibliography{Polariton}

\begin{thebibliography}{35}
\bibitem [{foo()}]{footnote}%
  \BibitemOpen
  \href@noop {} {}\bibinfo {note} {It can also include additional decay
  processes for the $|e\rangle$-state in a phenomenological way if
  needed}\BibitemShut {NoStop}%
\end{thebibliography}

\end{document}